# Kinetic equation approach to the problem of rectification in asymmetric molecular structures

Kamil Walczak [1]

Institute of Physics, Adam Mickiewicz University
Umultowska 85, 61-614 Poznań, Poland

Transport properties of asymmetric molecular structure are studied within the kinetic equation approach, taking into consideration the electron interaction in the self-consistent manner (SCF procedure). The device is made of a molecule (modeled as a quantum dot with discrete energy levels) which is asymmetrically coupled to the two metallic electrodes through the tunnel junctions. Electrical rectification follows from the combined effect of the geometric asymmetry in the molecular structure and simultaneously in the electrostatic potential spatial profile. Primarily the influence of broadening and charging effects on the charge and transport characteristics are investigated. Consequence of the broadening is to smooth the charge-voltage (N-V) and current voltage (I-V) functions, while the charging effect is rather responsible for the change of the slope of both N-V and I-V curves. Inclusion of the electron-electron repulsion results in significant enlargement of rectification ratio.

## I. Introduction

Molecular rectifier is one of the most important components for future applications in electronic circuits. Until now, much theoretical works have been devoted to the question of understanding the basic mechanisms of rectification in the complex acceptor-barrier-donor structures [1-3]. Recently, the use of Aviram-Ratner ansatz [1] to the problem of electron transport in such kind of devices led to the formula for the rectification current [4], which can be written in the simplified form:

$$I(V) = I_0 \left[ f(\varepsilon_L^A, \mu_1)(1 - f(\varepsilon_H^D, \mu_2)) \right]. \tag{1}$$

Prefactor $I_0 = e\, k_T$ is a simple product of an electronic charge unit and the acceptor-donor electron-transfer rate, and $f(\varepsilon, \mu)$ denotes the standard Fermi distribution function, respectively. Furthermore, there is a requirement for two electroactive molecular levels localized on particular subsystems of the molecule ($\varepsilon_L^A$ – LUMO of the acceptor and $\varepsilon_H^D$ – HOMO of the donor) and isolated from each other through insulating bridge. The above expression (eq.1) is valid only when we neglect all the broadening and charging effects. However, the dominant factor in the inducing rectification is a geometric asymmetry in the molecular junction [5].

A simple and general mechanism of molecular rectification, where a single electroactive unit is positioned asymmetrically with respect to the electrodes, was proposed recently by Kornilovitch *et al.* [6]. In their original work, the conjugated phenyl ring is isolated from the electrodes by two fully saturated tails of different lengths. Since most of the applied voltage drops on the longer insulating (σ-bonded) bridge, the conditions for resonant tunneling through the conducting level (localized on the ring) are achieved at very different voltages for two opposite polarities. The rectification ratio can be experimentally controlled by changing the lengths of the insulating barriers. It should be also noted that any asymmetric molecular junction produces asymmetric current at large enough voltages [7-9].

The major aim of this paper is to study the transport characteristics (charge-voltage and current-voltage) of an asymmetric molecular structure within the kinetic equation approach [10]. This method is almost equivalent to non-equilibrium Green function formalism and gives results, which could be qualitatively confirmed by experimental data. Molecular rectifier is modeled as a quantum dot (with discrete energy levels) asymmetrically coupled to two electrodes via the tunnel barriers. The simplicity of the model allows us to solve the problem in a self-consistent manner (SCF), taking into account the influence of the bias voltage on the charge redistribution along the device. The effect of the electron-electron repulsion is considered here within the Hartree-Fock approximation (HFA).

## II. Calculation details

Molecular rectifier with asymmetric tunneling barriers consists of two metallic electrodes, source ($M_1$) and drain ($M_2$), joined by a molecular wire. The wire is analyzed as a phenyl ring (with localized energy levels $\varepsilon_H$ and $\varepsilon_L$) separated from the electrodes by two insulating barriers (of the lengths $d_1$ and $d_2$) as shown in Fig.1. Sufficient insulation of the conducting molecular orbitals from both electrodes is an essential feature of the present rectification mechanism.

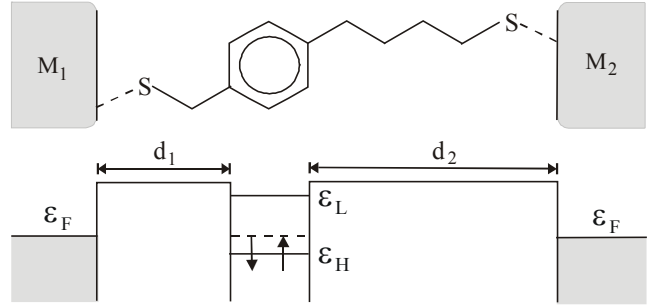

Fig.1 Schematic representation of the molecular rectifier and its energy level diagram.

In the frames of this work we do not want to repeat all the theoretical formulation of the problem (discussed elsewhere [10]) that explains the basic formulas, but we summarize the most essential results. Current flow is treated as a balancing act between two reservoirs of charge carriers (electrodes) with different agendas, which sends it into a non-equilibrium state (in the case of non-zero bias voltage). The number of electrons on the quantum dot under the non-equilibrium conditions is given through the relation:

$$N(V) = 2 \int_{-\infty}^{+\infty} d\varepsilon D(\varepsilon) \frac{\Gamma_1 f(\varepsilon,\mu_1) + \Gamma_2 f(\varepsilon,\mu_2)}{\Gamma_1 + \Gamma_2}. \qquad (2)$$

It should be noted that throughout this work we restrict our considerations to non-magnetic case (spin degenerate solution). It means that both spin orbitals feel the same self-consistent field and there is no need to distinguish two different contributions.

General expression for the current flowing through the device can be written as:

$$I(V) = \frac{2e}{\hbar} \int_{-\infty}^{+\infty} d\varepsilon D(\varepsilon) \frac{\Gamma_1 \Gamma_2}{\Gamma_1 + \Gamma_2} [f(\varepsilon, \mu_1) - f(\varepsilon, \mu_2)]. \quad (3)$$

Broadening that accompanies the coupling to the electrodes is taken into account through the Lorentzian density of states:

$$D(\varepsilon) = \frac{1}{2\pi} \left[ \frac{\Gamma_1 + \Gamma_2}{(\varepsilon - \varepsilon_H)^2 + \frac{1}{4}(\Gamma_1 + \Gamma_2)^2} \right], \quad (4)$$

where: $\Gamma_1$, $\Gamma_2$ are the dispersionless (by assumption) coupling parameters [11], $f(\varepsilon, \mu)$ denotes Fermi distribution function and $\varepsilon_H$ is energy of the molecular HOMO level, respectively. When the electroactive component of the molecule is closer to the electrode, the conducting level is essentially broadened, but in the case of larger distances – the broadening contribution is smaller ($\Gamma_1 > \Gamma_2$). The two electrochemical potentials are defined through the Fermi level $\varepsilon_F$ shifted by an amount of a half of bias voltage:

$$\mu_{1/2} = \varepsilon_F \pm \tfrac{1}{2} eV. \quad (5)$$

There are at least three parameters that determine the rectification properties of molecular device: (i) the energy difference between the conducting level and Fermi energy of the electrodes ($|\varepsilon_H - \varepsilon_F|$), (ii) the ratio of the voltage drops on the barriers proportional to their respective lengths ($\eta = d_1/d_2$) and (iii) the strength of the Coulomb interaction $U$. The change of the electronic structure of the phenyl ring is induced through the bias voltage:

$$\Delta\varepsilon_H = \left[ \frac{1-\eta}{2\eta+2} \right] eV. \quad (6)$$

To include the charging effects into the calculation scheme we also add a potential due to the change in the number of electrons from the equilibrium value:

$$U_{SCF} = U[N - 2f(\varepsilon_H, \varepsilon_F)], \quad (7)$$

and the conducting level is shifted as follows:

$$\varepsilon_H \rightarrow \varepsilon_H + \Delta\varepsilon_H + U_{SCF}. \quad (8)$$

Since that potential $U_{SCF}$ depends on the number of electrons $N$ and the number of electrons $N$ depends again on the potential $U_{SCF}$, we need to recalculate the potential for particular values of the bias voltage in the self-consistent (SCF) procedure (shown schematically in Fig. 2). Once the converged solution is obtained, the number of electrons $N$ (eq.2) and the current $I$ (eq.3) are calculated.

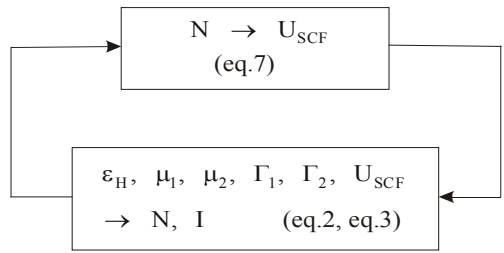

Fig.2 Illustration of the SCF procedure.

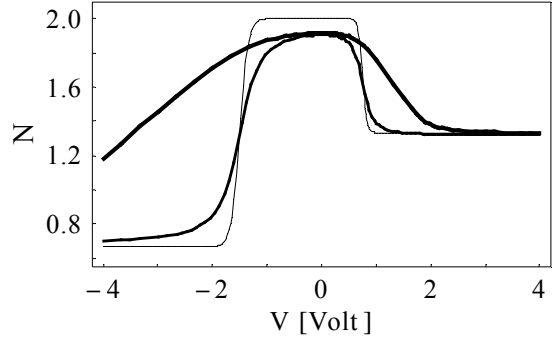

Fig.3 Number of electrons on the conducting level versus applied voltage. The width of the line increases with inclusion of analyzed effects (broadening and charging).

### III. Numerical results and discussion

Now we proceed to apply presented formalism to a single molecular HOMO level asymmetrically coupled to the two electrodes at a temperature $T = 300$ K ($\beta = 39/\text{eV}$). All the charge and transport characteristics are computed for the following values of the model parameters (energy given in eV): $\Gamma_1 = 0.1 = 2\Gamma_2$, $\varepsilon_H = -5.5$, $\varepsilon_F = -5.0$, $U = 1.0$, $\eta = 0.5$.

The number of electrons on the molecular level as a function of applied bias is shown in Fig.3. Ignoring the broadening, there are exactly two electrons occupying the HOMO level (corresponding to the two spin orientations – up and down, respectively) until the chemical potential crosses the conducting energy level to enable the electron to tunnel through the barrier into the electrode (this is accompanied by the current jump). That's why we observe the decrease of the charge on the molecule. Difference of an amount of transferred charge for both voltage polarities stems from the assumption of asymmetry in the coupling to the electrodes ($\Gamma_1 \neq \Gamma_2$).

The overall effect of the broadening is to redistribute charge carriers and move them partially to the contacts even at zero bias voltage (for $V = 0$ is $N = 1.916$). Moreover, the charge-voltage dependence is fairly smooth due to the broadening contribution. Charging enters the picture only at higher voltages and leads to further smoothing of the N-V curve and to changing the slope of the charge characteristic. Finally, the inclusion of the charging effectively tends to compensate the excess of the transferred charge induced by the coupling itself.

The current-voltage characteristics for analyzed type of the device is presented in Fig.4. The stepwise shape of the I-V curve in the absence of the broadening and charging effects is attributed to the discreteness of the conducting levels. Asymmetry of the conductance results from the asymmetric profile of the electrostatic potential across the junction [3,6], because in the case when bias is applied, the HOMO energy level becomes linear function of the bias voltage (eq.6). This is a directly consequence of the molecular orbital localization.

Rectification behavior of the current flowing through the device is even increased by broadening and charging effects. The evolution of the current with the voltage is continuous due to the broadening into our self-consistent calculations (just like in the most experimental data [12-15]) in spite of the distinct stepwise dependence typical for non-self-consistent results [4,16]. Inclusion of the electron-electron repulsion is found to reduce the magnitude of the backward current (enlarging the primary effect of rectification). Performing detailed calculations, it turned out that the slope of the I-V characteristics decreases with the increase of the U parameter.

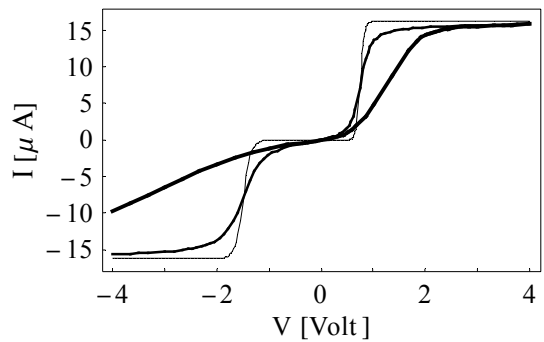

Fig.4  Comparison of the current-voltage characteristics for asymmetric molecular junction.
The width of the line increases with inclusion of analyzed effects (broadening
and charging, respectively).

## IV. A brief summary

In this paper we have studied transport properties of the device composed of two metallic electrodes connected by asymmetric molecule. The charge-voltage and current-voltage characteristics were calculated through the use of kinetic equation method. The electrical rectification follows from the combined effect of the geometric asymmetry in the molecular junction and the change of the electronic structure due to the bias voltage. Consequence of the broadening is to smooth the charge and current dependence, while the charging effect is rather associated with the change of the slope in both N-V and I-V curves. Inclusion of the electron-electron repulsion in the self-consistent calculation scheme results in significant enlargement of rectification.